\documentclass[12pt]{article}
\pdfoutput=1
\usepackage{amsmath, amssymb}    
\usepackage{graphicx}
\usepackage{dcolumn}
\usepackage{bm}
\usepackage{hyperref}
\usepackage{color}

\newcommand{\be}{\begin{equation} } 
\newcommand{\ee}{\end{equation} } 
\newcommand{\ba}{\begin{array} } 
\newcommand{\ea}{\end{array} } 
\newcommand{\bear}{\begin{eqnarray} } 
\newcommand{\eear}{\end{eqnarray} } 
\newcommand{\met}{ E_T\!\!\! \!\! \! \slash \;\, }

\title{
\vspace*{-3.4cm}
\begin{flushright}
\normalsize{ \small   Fermilab-PUB-19-643-T
  }
\end{flushright}
\vspace*{0.99cm}
\Large 
\textbf{
LHC probes of the 10 TeV scale
 }\vspace*{0.3cm}   
}

\author{{\bf  \normalsize 
Bogdan A. Dobrescu} 
\vspace{5mm}
\\
\normalsize\emph{Theoretical Physics Department, Fermilab, Batavia, IL 60510, USA}
}
\date{ \normalsize  December 30, 2019}

\begin{document}  
\setcounter{page}{0}  
\maketitle  

\begin{abstract} 
The usual range of new particle masses, up to a few TeV, searched for at the LHC may be substantially 
extended if ultraheavy diquark particles exist. 
A diquark scalar, $S_{uu}$, that interacts perturbatively  with  two up quarks may be as heavy as 10 TeV and would still 
produce  tens of spectacular events at the 14 TeV LHC. 
It is shown here that an ultraheavy $S_{uu}$ could be discovered through final states of very high energy in various channels, 
especially if the diquark can decay into other new heavy particles. 
Examples include cascade decays of $S_{uu}$ via a second scalar produced in pairs, 
which leads to two dijet resonances, or to more exotic signals with top quarks, Higgs bosons, electroweak bosons,
and high-$p_T$ jets.
Another possibility is that the diquark decays into a vectorlike quark of multi-TeV mass and a top or up quark.
Signal events include one or two highly boosted top quarks 
and a Higgs boson or a $Z$,  without counterparts containing top antiquarks.
Similarly, direct decays of the diquark into $tj$ or $tt$ with leptonic top decays involve only positively charged leptons.
\end{abstract} 
  
\thispagestyle{empty}  
  
\setcounter{page}{1}  
  
\vspace*{0.31cm}    
  
\tableofcontents
  
\baselineskip18pt   

\section{A diquark of mass near 10 TeV} 
\label{sec:Suu}

The analyses of the data sets from the LHC proton-proton collisions at a center-of-mass energy of 
13 TeV, by the ATLAS and CMS collaborations, have so far confirmed the predictions of the Standard Model. The LHC energy is soon going to be increased 
to 14 TeV. This 8\% increase might seem modest, but it may lead to substantially larger production cross sections of various processes
if there exist certain new particles of masses near the kinematic limit of the collider.

This effect is most intense in the case of the hypothetical particles called diquarks 
\cite{Mohapatra:2007af, Chen:2008hh, Bowes:1996xy}, which have the special property that can be produced in the $s$ channel by two incoming quarks.  
Given that the parton distribution functions (PDFs)  \cite{Harland-Lang:2014zoa,Dulat:2015mca,Ball:2017nwa}   
of the valence quarks in the proton are much larger than those of the antiquarks, 
the diquarks may be produced much more copiously 
than other particles beyond the Standard Model (SM), such as the $Z'$ or $W'$ bosons, that couple to a quark and an antiquark. 

Scalar diquarks can easily be included in renormalizable theories beyond the SM. 
Furthermore, these spin-0 states can be embedded in various theories, including some grand unified theories \cite{Mohapatra:1986uf}. 
By contrast, spin-1 diquarks \cite{Arik:2001bc} are not well-behaved in the UV unless 
some complicated gauge symmetries or perhaps some form of compositeness
are introduced. As such UV theories have not been identified yet,  it is preferably for now to focus on diquarks of spin 0. 

Among the scalar diquarks, the color sextet of electric charge 4/3, labelled $S_{uu}$, has the highest 
production cross section for a fixed coupling to quarks \cite{Dobrescu:2018psr}, because it is produced 
in the collision of two up quarks.  The PDF of the up quark is about six times larger 
than that of the down quark when the momentum fraction  of the 
parton is $x \approx 0.5$, which is necessary for producing a 10 TeV resonance in 14 TeV $pp$ collisions. 

To be concrete, the production cross section of an ultraheavy  $S_{uu}$ diquark scalar of mass $M_S = 10$ TeV and coupling $y_{uu} = 1$
is approximately 0.03 fb \cite{Dobrescu:2018psr} at the 14 TeV LHC. Depending on the decay modes, this may be large enough for discovery even 
with a fraction of the ultimate data set, which is expected to be 3000 fb$^{-1}$. For comparison, the production cross section at 13 TeV  is 
smaller by a factor of 10, so the energy upgrade is essential for probing a 10 TeV diquark. 

The main question to be addressed in order to search for this $S_{uu}$ particle is what are its decay modes. 
This paper discusses several different final states that can be generated by an ultraheavy diquark.
If the only new particle within the LHC reach is $S_{uu}$, then the constraints from flavor-changing processes 
indicate that the most likely final states are $tj$, $tt$ and $jj$, where $j$ is a hadronic jet, originating in this case from an up quark.
The events with top quarks are peculiar, as these are extremely boosted, and there are no similar events with top antiquarks (see Section \ref{sec:direct}).

If additional new particles are lighter than $S_{uu}$, then many more final states can be produced.
As there is no leading candidate for a theory describing the 10 TeV scale, one should be prepared for 
many possibilities. An interesting framework discussed recently \cite{Hill:2019ldq} is the so called ``scalar democracy", which assumes that at various mass scales there exist scalars associated which each fermion bilinear in the SM.
Even though \cite{Hill:2019ldq} focused especially on quark-antiquark states, it is interesting to consider the presence of all bilinear scalars.
Among those, the one that would likely be discovered first is the $S_{uu}$ discussed here, because it
can be produced with the largest cross section given a certain mass and coupling.
Nevertheless, the other scalars may also be present, and may modify the properties of $S_{uu}$. 

In particular, if a second heavy scalar, of electric charge 2/3 and labelled $S_{2/3}$, 
exists and is lighter than $M_S/2$, then $S_{uu}$ may have a cascade 
decay via two of those. When the lighter scalar behaves as a diquark  (see Section \ref{sec:S23diquark}), 
the LHC signal is an interesting 4-jet  final state that may be the origin of a striking 
event observed by CMS \cite{Sirunyan:2019vgj} at an invariant mass of 8 TeV.
Further studies in this direction are necessary, especially since the 
dedicated searches \cite{Sirunyan:2018rlj} for a  pair of dijet resonances have not yet required that all four jets form a peak.

An alternative model is that where the $S_{2/3}$ scalar has suppressed couplings to quarks, so that its dominant decays may be due 
to higher-dimensional operators. A simple UV completion  (see Section \ref{sec:S23exotic}), leading to various final states, such as 
$ (W^+  \bar t  \,  j)  (h^0 \bar b  j)  $ or $(h^0 \bar b  j) (Z \bar b  j)$, where $h^0$ is the SM Higgs boson and the parentheses denote 
resonances.

Another possibility is that a vectorlike quark, $\chi$, of electric charge 2/3 couples to $S_{uu}$, and there is no other new scalar.
For a vectorlike quark mass below $M_S/2$, the diquark can decay into two vectorlike quarks.
This scenario was analyzed in  \cite{Dobrescu:2018psr}, where the 
vectorlike quark was assumed to decay into two jets, providing an interpretation of the CMS $4j$ event at a mass of 8 TeV.
The dominant decays of the vectorlike quark  may instead be  the standard ones \cite{Han:2003wu}, 
$\chi \to  W^+ b $, $Z t$ and $ h^0 t $.
In that case the discovery of $S_{uu}$ and $\chi$ would occur in final states that are related to those of existing vectorlike quark 
searches \cite{Sirunyan:2019sza,Aaboud:2018wxv}, but include only positively-charged leptons (since they arise from $\chi \chi$ and not $\chi \bar \chi$) and are much more boosted.

The decay of $S_{uu}$ into a single vectorlike quark  (see Section \ref{sec:vectorlike}) and a SM quark would be even more interesting, 
as it would be sensitive to a heavier $\chi$. 
In particular, this would open an unique window for certain Composite Higgs models \cite{Cheng:2013qwa} where 
$\chi$ is expected to have a mass above 6 TeV.
Searches for events of the type  $ pp \to S_{uu} \to \chi \, u \to  (h^0 t) j $ or $( W^+ b) j $ would thus be well motivated. 
It is also worth exploring exotic decays of the vectorlike quark \cite{Dobrescu:2016pda}, which may lead to distinctive topologies,  
for example with 
two highly boosted top quarks (no $\bar t \, $) plus two $b$ jets arising from the $ pp \to S_{uu} \to  \chi t \to  (t b \bar b) t $ process.

\medskip
\section{Direct signatures of the $S_{uu}$ diquark} 
\label{sec:direct}   \setcounter{equation}{0}

The scalar diquark $S_{uu}$ transforms under the $SU(3)_c \times SU(2)_W \times U(1)_Y$ gauge group 
as $(6,1, +4/3)$. These gauge charges are the same as those of a pair of right-handed up-type quarks, 
which means that in the Lagrangian $S_{uu}$ couples 
only to pairs of right-handed up-type antiquarks. The flavor diagonal couplings of this type are given by 
\be
\dfrac{1}{2} \; K^n_{ij} \; S_{uu}^n \;  \left(   y_{uu} \,  \overline u_{R \, i} \,  u^c_{R \, j}
+  y_{cc} \,  \overline c_{R \, i} \,  c^c_{R \, j}
+   y_{tt} \,  \overline t_{R \, i} \,  t^c_{R \, j}    \right) \, + {\rm H.c.}  ~~~,
\label{eq:diquarkYuk}
\ee
where the $i$ and $j$ indices on the quark fields label the triplet color states ($i,j = 1,2,3$),
 the upper index $n$ on the scalar field labels the sextet color states ($n=1,..., 6$),
and the upper script $c$ on the quark fields denotes the charge conjugated field.
The parameters $y_{uu}$, $y_{cc}$ and $y_{tt}$ are Yukawa couplings,
while the coefficients $K^n_{ij}$ are products of $SU(3)_c$ generators \cite{Luhn:2007yr}, which can be written as follows: 
\be
K^{2\ell + 1}_{ij} = \delta_{i \ell} \delta_{j \ell}      \;\;\; \;\;  ,    \;\;\; \;\; 
K^{2\ell }_{ij} = \dfrac{1}{\sqrt{2} } \left(  \delta_{i \ell } \delta_{j \ell'} +  \delta_{i2} \delta_{j1}   \right)    \;\;\; \;\;  ,  
\label{eq:Kns}
\ee
where $\ell = 1,2,3$, and $\ell' = (\ell +1) \; {\rm mod} \, 3$. 

The interaction of $S_{uu}$ with charm quarks in (\ref{eq:diquarkYuk}) leads to $\Delta c =2$ processes, so it is constrained 
by measurements of  flavor-changing processes. 
The most constraining process in this case is $D^0-\bar D^0$ meson mixing, which sets an upper bound 
 on the $|y_{uu} y_{cc}|$ product. 
For a given $y_{uu}$ and diquark mass $M_{S}$, the limit on Yukawa coupling of the charm quark to $S_{uu}$ 
is \cite{Fortes:2013dba}
\be
|y_{cc}| <  \,  \dfrac{3.7 \times 10^{-4}}{ |y_{uu}| } \left(\dfrac{M_{S} }{10 \; {\rm TeV}} \right)^{\! 2} ~~. 
\ee
Since the production of an ultraheavy $S_{uu}$ at the LHC may occur only for
$|y_{uu}| \gtrsim 0.1$, the above limit implies that $ |y_{cc}|$ is a couple of orders of magnitude smaller than 
$|y_{uu}|$, and may be ignored in what follows.

The smallness of the charm coupling may be due to a flavor symmetry. A simple example is a 
global symmetry, such as a $U(1)$ or a $Z_n$ group, 
acting on the $c_R$ field. If the other quark fields are singlet under that symmetry, 
then not only $y_{cc}$ can be neglected, but also the  $S_{uu}$ interactions with $c \, u$ or $c \, t$ are suppressed.
Thus, the only flavor off-diagonal  coupling of $S_{uu}$ that could be sizable is to the up and top antiquarks ({\it i.e.}, 
the coupling that allows the $S_{uu} \to  u \, t$ process):
\be
y_{ut} \,   K^n_{ij} \; S_{uu}^n \;    \overline u_{R \, i} \,  t^c_{R \, j} \, + {\rm H.c.}  ~~~
\label{eq:utYuk}
\ee
The factor of 1/2 from the diagonal couplings (\ref{eq:diquarkYuk})  is not present here because
it is conventional to include it only for interactions that involve two fields of the same type.

If the only quark field charged under the global symmetry is $c_R$, then its
SM Yukawa coupling to the Higgs doublet $H$ is forbidden, and the charm quark mass $m_c$ will 
need a different origin than the other quark masses.
An example is the dimension-5 operator  $\phi_c  H \, \bar c_R Q_L^2 $, where $Q_L^2$ is the SM quark doublet
of the second generation, and $\phi_c$ is a scalar field with a VEV and the same charge under the global symmetry as $c_R$. 
This leads to a suppression of $m_c$ that may explain the hierarchy between the top and charm masses. 
However, the smallness of the up quark mass compared to $m_c$ would need a separate explanation.

Another example of charm mass generation is a model with two or more Higgs doublets, in which only one of the doublets 
($H_c$) carries the same charge under the global symmetry as $c_R$. Then the Yukawa interaction 
$H_c \, \bar c_R Q_L^2 $ gives $m_c$, while the top mass arises from the coupling to a different Higgs doublet
whose VEV is the dominant source of electroweak symmetry breaking. 

Both models  mentioned above as 
possible origins of $m_c$ would allow  $y_{ut}$ and $y_{tt}$ to be of the same order as  $y_{uu}$. 
Alternatively, the flavor symmetry may prevent the interaction of the diquark with both the charm and top quarks, 
as described in \cite{Dobrescu:2018psr}.  In that case only $u_R$ and $S_{uu}$ are charged under it, which would explain
why the up quark is so light compared to $m_c$. Such a symmetry would imply $y_{ut}, y_{tt} \ll y_{uu}$. 
Even then, the flavor symmetry may be only approximate, so  $y_{ut}$ or $y_{tt}$ might eventually be phenomenologically relevant. 

If there are no other new particles lighter than $S_{uu}$, then the only 2-body decay modes are into two up-type quarks.
There are three decay modes, $S_{uu} \to uu, ut, tt$, and their   branching fractions are:
\bear
B(S_{uu} \to uu) & \! = \! & \frac{|y_{uu} |^2}{| y_{uu} |^2 + 2 \, | y_{ut} |^2 + | y_{tt} |^2 }  ~~,
\nonumber \\ [2mm]
B(S_{uu} \to u \, t) &  \! = \! &  2 \, \frac{  |y_{ut} |^2}{| y_{uu} |^2  }  \; B(S_{uu} \to uu) ~~,
\label{eq:diquarkBR}
 \\ [2mm]
B(S_{uu} \to t \, t)  &  \! = \! & \frac{  |y_{tt} |^2}{| y_{uu} |^2  } \; B(S_{uu} \to uu)  ~~,
\nonumber
\eear
where the top mass has been neglected since $m_t/M_S \ll 1$. The total width of $S_{uu}$ into quark pairs at leading order is 
\be
 \Gamma (S_{uu} )_{qq'}  =  \left( | y_{uu} |^2 + 2 \, | y_{ut} |^2 + | y_{tt} |^2 \right)  \dfrac{M_S }{32 \pi}   ~~.
 \label{eq:uuWidth}
\ee
Note that $S_{uu}$ is a narrow state, with a width-to-mass ratio below 4\% for $y_{qq'} \lesssim 1$.

The decay $S_{uu} \to u u $ is unavoidable if  $S_{uu}$ is produced at a large rate, and  leads to a dijet resonance signal.
Even though the QCD background is typically large for dijet searches, a narrow resonance at a mass as large as 10 TeV 
can be discovered with only a few events at the LHC  \cite{Dobrescu:2018psr}.

In the $S_{uu} \to u \, t $ and $t \, t$ channels, the top quarks are so highly boosted that when decaying  hadronically
they would look like a narrow jet without a 3-prong substructure. Thus, the dominant signal in these channels is 
still a  dijet resonance. In the case of leptonic decays $t \to W^+ b \to \ell^+ b + \met $, the lepton is not isolated from the 
$b$ jet due to top boosting. Interestingly the signal consists entirely of 
positively charged leptons \cite{Mohapatra:2007af}. The reason is that the production cross section for $S_{uu}$ is larger by about two orders of magnitude 
than that for its antiparticle because the PDF for $\overline u$ is highly suppressed compared to
that for $u$ at large $x$. 
Thus, $pp \to S_{uu} \to t \, t $ gives rise to two top quarks, as opposed to a top-antitop pair as assumed in usual 
$t\bar t$ resonance searches. 
Even more unusual is the $pp \to S_{uu} \to u \, t $ signal, where there is a single top quark 
produced in association with a very high-$p_T$ jet, and no similar events with a  $\bar t \,$ would be observed.

\bigskip
\section{Diquark decays into two $S_{2/3}$ scalars}
\setcounter{equation}{0} \label{sec:2scalars}

As discussed in Section \ref{sec:Suu}, if a scalar like the $S_{uu}$ diquark exists, it is reasonable to expect 
that other new scalars are also present, and have masses of roughly the same order of magnitude as the mass $M_S$ of 
$S_{uu}$.

\begin{figure}[t]
\hspace*{-1mm} \includegraphics[width=0.49\textwidth]{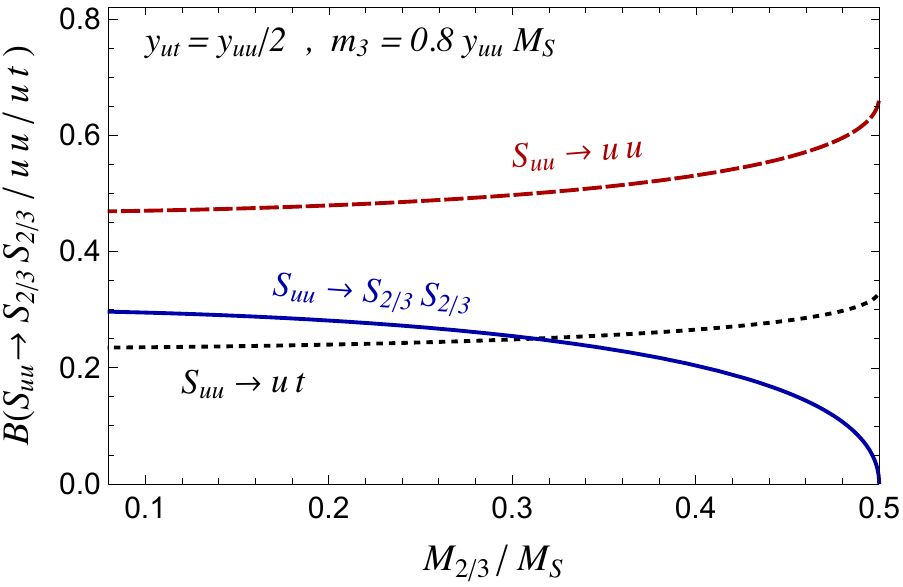} \hspace*{1mm}  
\includegraphics[width=0.49\textwidth]{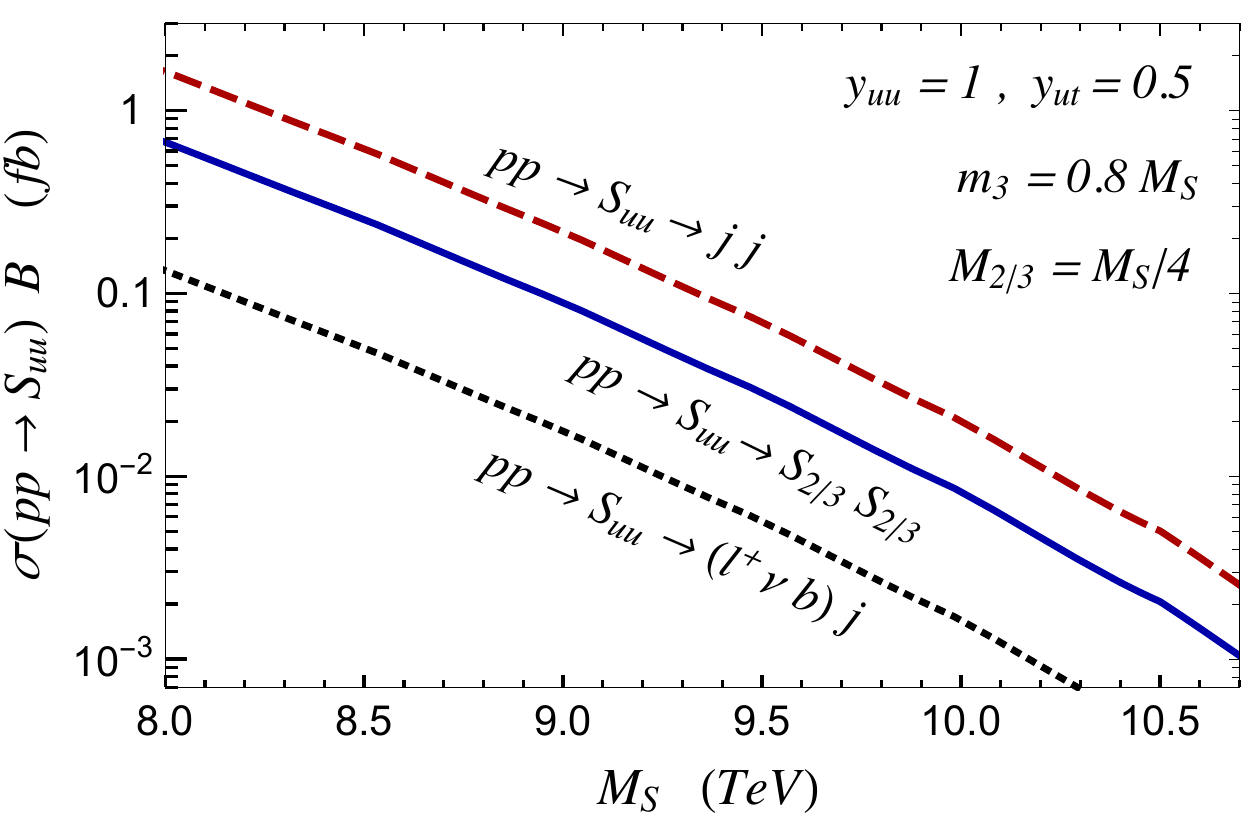}  
\caption{{\it Left panel:} 
Branching fractions of the $S_{uu}$ scalar diquark as a function of the mass ratio of the 
$S_{2/3}$ scalar and $S_{uu}$. The trilinear scalar coupling is taken to be 
$m_3 = 0.8 y_{uu} M_S$, the  Yukawa coupling to $u t$ is fixed at $y_{ut} = y_{uu}/2$, and  $y_{tt}$ is neglected. 
 {\it Right panel:}  Production cross section at the 14 TeV  LHC times branching fractions for $S_{uu}$
 as a function of its mass $M_S$. The Yukawa 
coupling to up quarks is fixed at $y_{uu} = 1$ (the total cross section is proportional to $y_{uu}^2$),
and the ratio of scalar masses is fixed at $M_{2/3}/M_S = 1/4$. The dijet rate (dashed red line) is the sum 
of the $pp \to S_{uu}\! \to uu$ and $pp \to S_{uu}\to t u$ processes with the top quark decaying hadronically,
while the rate for $pp \to S_{uu} \!  \to t u \to (\ell^+ \nu b) j$ (dotted black line) includes only the $W \to e^+ \nu, \, \mu^+ \nu$ decays. 
\vspace*{6mm}
} 
\label{fig:BRofSuu}
\end{figure}

Consider a model where in addition to the $S_{uu}$ diquark there is a color-triplet scalar $S_{2/3}$ of electric charge $2/3$.
More precisely, $S_{2/3}$ transforms as $(3,1, +2/3)$ under the $SU(3)_c \times SU(2)_W \times U(1)_Y$ gauge group.
Then a trilinear scalar coupling involving the diquark and two charge-conjugate $S_{2/3}$ fields is allowed by the SM  gauge symmetry:
\be
\dfrac{m_3}{2} \;  K^n_{ij} \, \; S_{uu}^n \;  S_{2/3}^{\dagger \, i} \, S_{2/3}^{\dagger j} + {\rm H.c.}  ~~~,
\label{eq:3diquark}
\ee
where $m_3$ is a mass parameter and $K^n_{ij}$ is given in (\ref{eq:Kns}).
If the mass of $S_{2/3}$ satisfies  $M_{2/3} < M_S/2$, then the above interaction induces the decay of $S_{uu}$
into two $S_{2/3}$ scalars with the following leading order width     
\be
\Gamma (S_{uu} \to S_{2/3} \, S_{2/3} )  =   \dfrac{m_3^2}{32 \pi \, M_S  } 
\left( 1 - \dfrac{4 M_{2/3}^2 }{M_{S}^2 } \right)^{\! 1/2}  ~~.
\label{eq:SSSWidth}
\ee
For $m_3$ of the same order as the diquark mass, this decay has a sizable branching fraction,
as can be seen in  Figure \ref{fig:BRofSuu} (solid blue line in the left panel).
The values used there for the coupling ratios are $y_{ut}/y_{uu} = 0.5$, $y_{tt}/y_{uu} \ll 1$, and  $m_3/y_{uu} = 0.8 M_S$. 

The production cross section for $S_{uu}$ has been computed in  \cite{Dobrescu:2018psr}.  
The PDFs at very large $x \gtrsim 0.3$, which is the region of interest for ultraheavy resonances, 
have substantial uncertainties. As an example, for the production of an  $S_{uu}$ with mass of 10 TeV at the 14 TeV LHC,
the PDF set MMHT \cite{Harland-Lang:2014zoa} gives a next-to-leading order (NLO) cross section smaller by 12\% than that 
obtained with the CT14 set  \cite{Dulat:2015mca},
and the $1\sigma$ uncertainties given by the CT14 set are approximately 15\% \cite{Dobrescu:2018psr}.

The central value obtained with the CT14 set at NLO is  $\sigma (pp \to S_{uu}) \approx 3.1 \times 10^{-2}$ fb $y_{uu}^2 $ for $M_S = 10$ TeV.
Thus, for that mass and $y_{uu}=1$, the LHC would produce about 90 events originating from the decays of $S_{uu}$.
Independently of the $S_{uu}$ decay mode, each of these events would be spectacular and almost background free
due to the huge $p_T$'s of the final state objects. The efficiency of the selection cuts would depend on how the  searches
are performed, but in principle the majority of these events may pass the selection criteria. 
 
The right panel of Figure \ref{fig:BRofSuu} shows the products of $\sigma (pp \to S_{uu})$ at $\sqrt{s} = 14$ TeV  
and the branching fractions for $S_{uu}$ as a function of its mass. 
The values of the couplings used there are  $y_{uu} = 1$, $y_{ut} = 0.5$, $y_{tt} \ll 1$, and $m_3 = 0.8 M_S$;
the $S_{2/3}$ mass is fixed at $M_{2/3} = M_S/4$. 
The pair production of $S_{2/3}$ is shown by the solid blue line. Since the top quark is extremely boosted in the $pp \to S_{uu}\to ut$ process,
its hadronic decay looks like a narrow jet, so the rate for this process is added with the one for $pp \to S_{uu}\to uu$  to obtain the dijet rate,
shown by the dashed red line.  The cross section for $pp \to S_{uu} \to t  u \to (\ell^+ \nu b) j $, where $\ell = e, \mu$,
is given by the dotted black line.
It is worth reiterating that 
the $S_{uu}$ discussed here is a narrow resonance: its total width  $ \Gamma_S$ is the sum of   
$ \Gamma (S_{uu} )_{qq'}$  and $\Gamma (S_{uu} \to S_{2/3} \, S_{2/3} ) $ given in (\ref{eq:uuWidth}) and (\ref{eq:SSSWidth}), so 
the width-to-mass ratio is $ \Gamma_S/M_S \approx 1.8 \%$ at leading order  for 
 the values of the parameters used  in Figure \ref{fig:BRofSuu}. 

Since the $S_{uu}$ diquark has a coupling of order one to two up antiquarks, shown in (\ref{eq:diquarkYuk}), its baryon number must be 
$B=2/3$
(using the normalization where all SM quarks have baryon number 1/3),
while its lepton number is $L = 0$.
As the trilinear coupling (\ref{eq:3diquark}) is also assumed to be large, the $S_{2/3}$ scalar must carry $B = 1/3$ and $L = 0$.
Even though the global symmetries associated with baryon or lepton numbers may be broken, it is useful to keep track of them as
there are some experimental limits on how large these symmetry breaking effects
may be. 

\subsection{$S_{2/3}$ as a diquark}  \label{sec:S23diquark}

The gauge quantum numbers of $S_{2/3}$ allow it to  
couple to two down-type quarks, so $S_{2/3}$ may behave as a diquark. However, those couplings are totally 
antisymmetric in color indices, so they vanish unless the two quarks have different flavor. 
Thus, the most general renormalizable terms in the Lagrangian that couple the $S_{2/3}$ scalar to SM fermions are
\be
\epsilon_{ijk} \; S_{2/3}^k \;  \left(   y_{ds} \,  \overline d_{R \, i}^{\; c} \,  s_{R \, j} + y_{db} \,  \overline d_{R \, i}^{\; c}  \,  b_{R \, j}
+ y_{sb} \,  \overline s_{R \, i}^{\; c}  \,  b_{R \, j}  \right)  ~~,
\label{eq:ds}
\ee
where $i,j,k$ are the color indices of the triplet fields.
The Yukawa couplings $y_{ds}$, $y_{db} $ and $y_{sb} $ are free parameters which determine the  widths 
for the three decay modes: $S_{2/3} \to \bar d \bar s$, $\bar d \, \bar b$ and $\bar s \bar b$.
All these decays are into two hadronic jets, with two of them including a single $b$ jet. 

\begin{figure}[t]
\hspace*{-1mm} \includegraphics[width=0.475\textwidth]{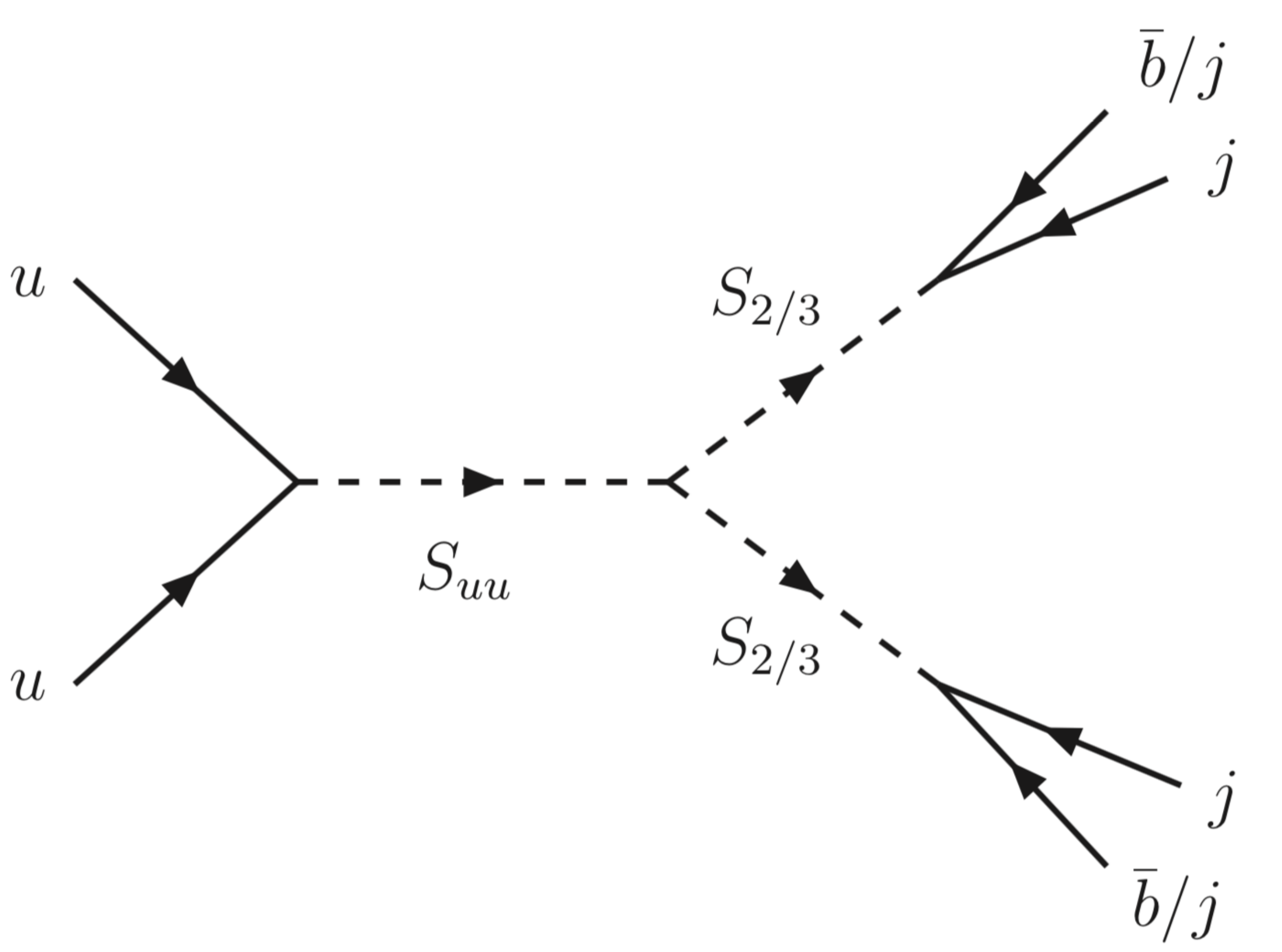} \hspace*{7mm}  
\includegraphics[width=0.475\textwidth]{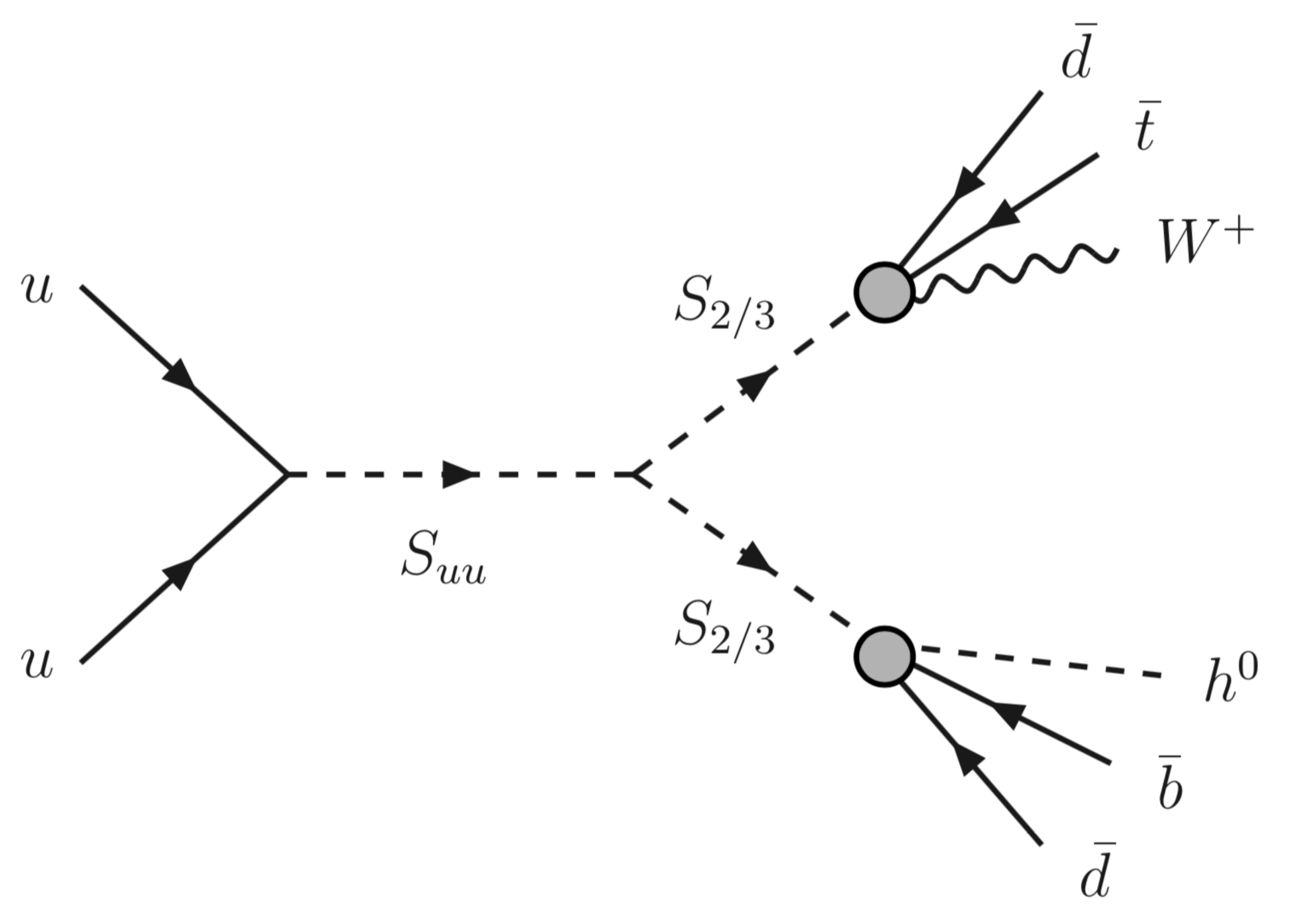}  
\caption{LHC processes with 
production of the $S_{uu}$ diquark followed by cascade decays
via a pair of $S_{2/3}$ scalars, in two 
models.
{\it Left diagram:}  $S_{2/3}$ couples to down-type quarks as in (\ref{eq:ds}), resulting in two dijet resonances of equal mass.
 {\it Right diagram:}  $S_{2/3}$ undergoes 3-body decays 
 induced by a vectorlike quark, 
 which has been integrated out and couples as in (\ref{eq:dbp}).
\vspace*{2mm}
} 
\label{fig:DiagramsSSS}
\end{figure}

The main process of interest at the LHC is then $s$-channel production of the ultraheavy
$S_{uu}$ diquark, followed by its decay into two $S_{2/3}$ scalars, with each of them forming a dijet 
resonance (see the first diagram of Figure~\ref{fig:DiagramsSSS}) of mass
equal to $M_{2/3}$. The invariant mass of the four jets is given by the $S_{uu}$ mass, $M_S$.
This process,
\be
pp \to S_{uu} \to S_{2/3} S_{2/3} \to (jj)(jj)  ~~,
\label{eq:4j}
\ee
offers a simple interpretation for the event reported by CMS \cite{Sirunyan:2019vgj}, in which two pairs of jets, each of mass near 
$M_{2/3} \approx 1.8$ TeV, form an invariant mass of 8 TeV. 
A different interpretation, also based on a diquark with mass of 8 TeV, was given in \cite{Dobrescu:2018psr}, where it is shown 
that the production rate is large enough when the coupling $y_{uu}$ is around 0.3.
The QCD background that includes four jets of very high $p_T$ with the appropriate topology 
is low, with the probability of producing the 8 TeV event below $10^{-4}$. Nevertheless,
a fluctuation of the background may be responsible for the observed event, especially since there is a large look-elsewhere effect.
The model with the $S_{uu}$ and $S_{2/3}$ diquarks is interesting independently of whether  other similar events at 8 TeV will be observed,
so in what follows $M_S$ will be taken to be a free parameter near the 10 TeV scale.

If $ y_{ds}$ is much larger than the other two Yukawa couplings, then the predominant decay channel of $S_{2/3}$ is
$ \bar d \bar s$, so the process (\ref{eq:4j}) does not include any $b$ jets. 
In the opposite limit, two of the four jets are $b$ jets. 
Even more remarkable  is the final state where only one of the four jets is a $b$ jet, which is likely to occur 
when $ y_{ds}$ is comparable to $y_{db} $ or $y_{sb} $, so that the branching fractions for $S_{2/3} \to jj$ and $jb$ are near 1/2.

The interactions (\ref{eq:ds}) break baryon number by an amount $\Delta B = 1$. As a result, at scales below the $S_{2/3}$ mass,
the trilinear scalar coupling
(\ref{eq:3diquark}) leads to $\Delta B = 2$ operators of the type
\be
\frac{y_{uu}^*  m_3}{4 M^2_S M^4_{2/3}} \; \left( y_{ds}^2\;  (u u) (d s) (d s)  +  y_{db}^2\;  (u u)( d b)( d b)  +  2 y_{db}y_{sb}\;  (u u) (d  b)(s b) + ... \right)   ~~. 
\label{eq:baryon}
\ee
Searches for neutron-antineutron oscillations do not directly constrain these types of operators, as they 
always involve second or third generation quarks.
The first operator, though, is constrained by the process $pp \to K^+ K^+$, as was pointed out in \cite{Arnold:2012sd}.
A very strong limit on this process was set by the Super-Kamiokande collaboration  \cite{Litos:2014fxa}.
However, the dimension-9 operator $u u d s d s$, which is obtained from the first diagram of Figure~(\ref{fig:DiagramsSSS})
after the scalars are integrated out,  is suppressed by 5 powers of the 10 TeV scale. 
A comparison of an operator obtained in \cite{Goity:1994dq}, which is used by \cite{Litos:2014fxa} as a benchmark,
with the first operator of (\ref{eq:baryon}) gives an upper limit on $|y_{ds}| $ of about $10^{-5}$ for
$m_3 \approx M_S \approx 10$ TeV, $M_{2/3} \approx M_S/4$, and $y_{uu} \approx 1$.
This limit on $y_{ds}$, while significant, is not affecting the collider phenomenology discussed so far.

Only when $y_{ds}$, $y_{db} $ and $y_{sb} $ are all below roughly $10^{-7}$
the decay length of $S_{2/3}$ becomes macroscopic, and the 
LHC signatures are dramatically affected.
The process $pp \to S_{uu} \to S_{2/3} S_{2/3}$ then leads to two highly-ionizing tracks, each ending at
a displaced vertex where two hadronic jets originate. 

Although the interactions (\ref{eq:ds}) are flavor violating, 
there are no contributions from tree-level  $S_{2/3}$ exchange to neutral kaon mixing, or to $\Delta b = 2$ processes.
There are however tree-level contributions to the $b \to s  \bar d  d$ and $b \to s \bar s d$ transitions. These impose 
some mild constraints on the products of Yukawa couplings. As mentioned above,
from the point of view of collider phenomenology,
those constraints from flavor processes may be ignored because the overall normalization of the Yukawa couplings can be extremely small
without changing the decay modes of $S_{2/3}$.

 \subsection{Exotic decays of $S_{2/3}$} \label{sec:S23exotic}  

Another possibility is that the renormalizable couplings of $S_{2/3}$  are highly suppressed, but some couplings 
due to higher-dimensional operators are large enough for the scalar to decay promptly, or 
inside the detector but with a displaced vertex. 
There are several possibilities for the decays of $S_{2/3}$. Here only
a few higher-dimensional operators are discussed, which originate from a simple UV completion.

Let us assume that a vectorlike quark $b'$ of electric charge $-1/3$ exists and it is heavier than $S_{2/3}$.
The mixing between $b'$  and $b$ leads to interactions of $b'$ with $ W^- t$, $ Z b$, and $ h^0  b$. In addition, 
$b'$ may couple to $S_{2/3}$ and a SM down-type quark. Let us assume for simplicity that only one of the three 
couplings of this type is large, corresponding to the Lagrangian term
\be
y_{db'} \, \epsilon_{ijk} \; S_{2/3}^k \;   \overline d_{R \, i}^{\; c}  \,  b'_{R \, j}  ~~.
\label{eq:dbp}
\ee
Integrating out $b'$  gives three dimension-5 operators, involving two SM quarks, an electroweak boson, and a $S_{2/3}$ scalar.
The $S_{2/3}$ scalar can then decay (effectively via an off-shell $b'$ antiquark) with three leading channels available:
$S_{2/3} \to \bar b^{\prime *} \bar d \to h^0  \bar b  \bar d$ or $Z \bar b \bar d$ or $W^+ \bar t \,  \bar d$.

There are thus six cascade decays of the $S_{uu}$ diquark:
\be
pp \to 
S_{uu} \to S_{2/3} S_{2/3} \to   \left\{ \ba{l}  
 (W^+  \bar t  \,  j)  (h^0 \bar b  j)  \;  ,   \;\;  (W^+ \bar t \,  j)  (Z \bar b  j)   \; ,   \;\;    (W^+ \bar t \,  j)  (W^+ \bar t  \, j)  ~, 
 \\  [2mm] 
\;  (h^0 \bar b  j)  ( Z \bar b  j)   \;\;\; ,   \;\;\;     (h^0 \bar b j )  (h^0 \bar b j )     \;\;\; ,   \;\;\;     (Z \bar b  j )  (Z \bar b  j)  ~~. 
\ea \right.  
\label{eq:dbp-final}
\ee
The first three, which involve one or two  top antiquarks, have branching fractions of 1/4.
The branching fraction of $S_{uu}  \to (h^0 \bar b  j)  (Z \bar b  j)$  is 1/8,  while the last two modes have each a branching fraction of 1/16.
The first process of (\ref{eq:dbp-final}) is shown at partonic level in the 
second diagram of Figure~\ref{fig:DiagramsSSS}.

While these final states have some similarities with the ones studied in the searches for vectorlike quarks,
there are also several important differences. The objects involved in these final states are more boosted, since they originate from an ultraheavy 
$s$ channel resonance. Second, they all involve two additional high-$p_T$ jets. Third, the $W$ bosons that do not originate from top 
decays have always charge +. 

The model described here is sufficiently simple to warrant dedicated searches in the above final states.
Due to the boosted topology, a positive signal would allow the reconstruction of the $S_{uu}$ and $S_{2/3}$ resonances, 
as well as an indirect determination of the mass and couplings of the vectorlike quark.

\section{Diquark decays into vectorlike quarks}
\setcounter{equation}{0}   \label{sec:vectorlike}

In order to assess the range of signatures that the $S_{uu}$ diquark may generate, it is interesting to 
consider now a different model, with the only particles beyond the SM being $S_{uu}$  and a 
vectorlike quark $\chi$, which has the same SM gauge charges as $u_R$, namely $(3, 1, +2/3)$. 

This model was analyzed in  \cite{Dobrescu:2018psr} with a couple of restrictive assumptions, namely 
that $\chi$ couples to $S_{uu}$ only in pairs, and that the $\chi$ decays into a light quark and a gluon.
Here I study more general possibilities consistent with the SM gauge symmetries.

The most general Yukawa interactions of $\chi$  and $S_{uu}$, excluding the charm quark as 
suggested by the flavor constraints discussed in Section \ref{sec:direct}, are given by
\be
K^n_{ij} \; S_{uu}^n \;  \left(  y_{u \chi} \, \overline u_{R \, i} \,  \chi^c_{R \, j}  +  y_{t \chi} \, \overline t_{R \, i} \,  \chi^c_{R \, j} 
+ \frac{y_{\chi\chi_R}}{2}   \, \overline \chi_{R \, i} \,  \chi^c_{R \, j}   +  \frac{y_{\chi\chi_L}}{2}  \, \overline \chi_{L \, i} \,  \chi^c_{L \, j} 
 \, \right)  
+ {\rm H.c.}  ~~~
\label{eq:diquarkChi}
\ee
These are similar to the Yukawa interactions of  $S_{uu}$ shown in (\ref{eq:utYuk}) and (\ref{eq:diquarkYuk}),
with the main difference that there are flavor-diagonal couplings for both the right- and  left-handed components of $\chi$. 

If $\chi$ has a mass $m_\chi < M_S$ (ignoring again the top mass),  
then the $S_{uu} \to u \chi$ and $S_{uu} \to t \chi$ decays are kinematically allowed. 
Furthermore, if $m_\chi < M_S/2$, then  $S_{uu} \to \chi \chi$  also opens up.
The widths at leading order  for these decays are 
\bear
&& \Gamma (S_{uu} \to u \, \chi)  =  
 \frac{  |y_{u \chi} |^2}{| y_{t\chi} |^2  } \,  \Gamma (S_{uu} \to t \, \chi)  =  \dfrac{M_S   }{16 \pi}  \, |y_{u\chi}|^2 \;   \left( 1 - \dfrac{ m_\chi^2 }{M_{S}^2 } \right)^{\! 2}  ~~,
\nonumber \\ [-2mm]
\\ [-2mm]
&& \Gamma (S_{uu} \to \chi \, \chi)  =   \dfrac{M_S   }{32 \pi} \left( |y_{\chi\chi_R}|^2 + |y_{\chi\chi_L}|^2 \right)  \;   \left( 1 - \dfrac{2 m_\chi^2 }{M_{S}^2 } \right)\left( 1 - \dfrac{4 m_\chi^2 }{M_{S}^2 } \right)^{\! 1/2}  ~~.
\nonumber
\label{eq:WidthSchi}
\eear

If there is mass mixing between $\chi$ and the top quark, then 
$\chi$ has three main decays: $\chi \to  W^+ b$ with a branching fraction of 1/2, and 
$\chi \to h^0 t $ or $Z t $, each with a branching fraction of 1/4  \cite{Han:2003wu}.
These can be seen as the ``standard" decay modes, and are assumed in the searches for vectorlike quarks of charge 2/3.

The decay of $S_{uu}$ via a single $\chi$ followed by the standard $\chi$ decay can lead to the following final states:
\bear
&& pp \to S_{uu} \to \chi u \to  (W^+ b) j \; , \;   (h^0 t ) j   \; , \;  (Z t) j
\nonumber \\ [-2mm]
\label{eq:singleChi}
\\ [-2mm]
&& pp \to S_{uu} \to \chi t \to  (W^+ b) t  \; , \;   (h^0 t) t \; , \;  (Z t) t
\nonumber 
\eear
Only top quarks and no antiquarks are produced in these cascade decays, so all the leptons produced in $W$ decays 
(unless they originate from Higgs decays) have positive electric charge.
The very high boosting of the top quarks and electroweak bosons make all these final states to exhibit  a  3-prong topology, with 
each the three objects having a $p_T$ of a few TeV or even higher.
These features make the channels (\ref{eq:singleChi}) almost background free. For $M_S = 10$ TeV, 
it is possible to be sensitive to vectorlike quark masses as large as 8 TeV, which is about 4 times higher than the current limits
based on QCD production of $\chi\bar \chi$.

The $S_{uu} \to  \chi\chi$  decay is also interesting, leading to different six final states:
\be
(W^+ b) (W^+ b) \; , \;   (W^+ b)(h^0 t )    \; , \;  (W^+ b)(Z t)     \; , \; 
(h^0 t ) (h^0 t )   \; , \;   (h^0 t ) (Z t)     \; , \;     (Z t) (Z t)  ~~.
\ee
Again, the vast majority of the events of these types which include leptons have only positively charged ones. 
Given the higher multiplicity, the boosting may be lower than in events arising from cascades with a single $\chi$.

If the mass mixing between $\chi$ and $t$ is negligible, then the vectorlike quark
may have exotic decays \cite{Dobrescu:2016pda} induced by particles heavier than $\chi$.
In this model, such multi-body decays don't even require other new particles, because they can be  
induced by the $S_{uu}$ diquark.  
Consider for example the following 3-body decay of the vectorlike quark via an off-shell diquark:
\be
\chi \to S_{uu}^* \bar u \to  t \, t \, \bar u  ~~.
\ee
The amplitude for this is proportional to $y_{u \chi}$ and $ y_{tt} $. 
If these Yukawa couplings are not supressed compared to $y_{t \chi}$ and $ y_{uu} $, respectively,
then the above 3-body channel may be the dominant decay mode of $\chi$.
The diquark production followed by its $\chi\chi$ decay then leads to the process
\be
pp \to S_{uu} \to \chi\chi \to (t \, t \, j)(t \, t \, j)  ~~.
\label{eq:4t}
\ee
This final state with four top quarks and two high-$p_T$ jets 
has no counterpart with top antiquarks. 
Thus the events with leptons again involve only positively charged ones.
Even striking signals with $\ell^+\ell^+\ell^+$ plus 6 or more high-$p_T$ jets  
could be observed, since the combined branching fraction
for the four $W^+$ bosons from the top quarks to give three leptons is about 6\%. 

\section{Outlook} 
 \label{sec:conclusions}    \setcounter{equation}{0} 

The LHC exploration of the TeV scale is soon going to intensify, with the plan of increasing the center of mass energy to 14 TeV 
and accumulating 20 times more data than currently available. Furthermore, many upgrades of the detectors, and continuous 
improvements of the experimental techniques will make the analyses of the data more efficient.

The renormalizable models studied here show that the LHC data may be used to probe directly not only the TeV scale but also 
certain scenarios for physics at the 10 TeV scale. Specifically, the ATLAS and CMS collaborations will be able to search for 
narrow resonances with masses around 10 TeV, by looking in a variety of final states for highly boosted top quarks, 
Higgs bosons, $W$ and $Z$ bosons and jets of very high $p_T$.  

The ideal target of such searches would be a diquark scalar, $S_{uu}$, which is a color sextet and carries electric charge 4/3,
because its production rate is much larger than for other narrow resonances that are predicted in well-defined 
renormalizable models. 
Furthermore, $S_{uu}$ may be associated with other exotic particles, whose discovery would take us closer to 
revealing the underlying theory for physics beyond the SM.

Two classes of renormalizable models have been considered. In the first one, besides $S_{uu}$ there is a second scalar,
$S_{2/3}$, which may be produced in pairs through the decay of $S_{uu}$. The most  straightforward signature 
is a 4-jet resonance at the $S_{uu}$ mass, with the jets forming two dijet resonances of equal mass (set by $S_{2/3}$).
This model provides a viable alternative to the possible origins identified in  \cite{Dobrescu:2018psr} of the 
striking 4-jet event at a mass of 8 TeV reported by the CMS collaboration \cite{Sirunyan:2019vgj}. 

If the coupling of $S_{2/3}$ to SM quarks is suppressed, then it could decay via higher-dimensional operators induced by 
integrating out a very heavy vectorlike quark of electric charge $-1/3$. In that case, the $p p \to   S_{uu} \to S_{2/3} S_{2/3}$ process
leads to several final states, listed in (\ref{eq:dbp-final}). These may arise from prompt decays, or from displaced vertices,
depending on how large are the effects induced by the vectorlike quark. 

In the second class of models discussed here, there are only two particles beyond the SM: the $S_{uu}$ diquark and a 
vectorlike quark $\chi$ carrying the same gauge charges as the right-handed top quark.
This is similar with one of the models analyzed in \cite{Dobrescu:2018psr}, but a more general set of interactions are 
considered here.  Among these, the coupling of $S_{uu}$ to a single $\chi$ and a SM quark, and the mixing of $\chi$ and $t$,
lead to the final states listed in (\ref{eq:singleChi}). Other interesting final states, such as 4 top quarks (no anti-top)
and two jets, may arise through the pair production of $\chi$ in $S_{uu}$ decays, followed by the 3-body decay of each $\chi$
via an off-shell $S_{uu}$. 

All the events produced through the cascade decays of the diquark through one or two $\chi$'s have the peculiar feature that they involve only 
positively charged $W$'s (with a few exceptions, such as  events that include a Higgs boson decaying to $WW^*$).
The reason is that the  $S_{uu}$ production is two orders of magnitude larger than that for its antiparticle.
Even when the only particle beyond the SM is $S_{uu}$, its direct decays to $t u$ or $t t$ with leptonic $W$ decays
would contain only  positively charged leptons.
Events with very high-$p_T$ leptons, only of + charge and often inside the cone of a jet, would be a hallmark 
of an ultraheavy diquark.

The studies presented here deal with only some of the cleanest signals due to an ultraheavy boson at the LHC. 
Ultraheavy particles other than the $S_{uu}$ diquark, such as diquarks with different quantum numbers, colorons, $W'$
or $Z'$ bosons, may also open a window into physics at the 10 TeV scale, and should be intensely searched for by the experimental collaborations.

\bigskip\bigskip\bigskip\bigskip\bigskip


{\bf Acknowledgments:} \  I would like to thank Robert Harris and Josh Isaacson for insightful conversations.
This work was supported by Fermi Research Alliance, LLC, under Contract No. DE-AC02-07CH11359 with the U.S.
Department of Energy, Office of Science, Office of High Energy Physics.



\begin{thebibliography}{99} 
 
\bibitem{Mohapatra:2007af} 
  R.~N.~Mohapatra, N.~Okada and H.~B.~Yu,
  ``Diquark Higgs at LHC,''   
  Phys.\ Rev.\ D {\bf 77}, 011701 (2008)
  [arXiv:0709.1486]. \\
  J.~M.~Arnold, M.~Pospelov, M.~Trott and M.~B.~Wise,
  ``Scalar representations and Minimal Flavor Violation,''
  JHEP {\bf 1001}, 073 (2010)
  [arXiv:0911.2225].  \\
  E.~L.~Berger, Q.~H.~Cao, C.~R.~Chen, G.~Shaughnessy and H.~Zhang,
  ``Color sextet scalars in early LHC experiments,''
  Phys.\ Rev.\ Lett.\  {\bf 105}, 181802 (2010)
  [arXiv:1005.2622 [hep-ph]]. 

\bibitem{Chen:2008hh} 
  C.~R.~Chen, W.~Klemm, V.~Rentala and K.~Wang,
 ``Color Sextet Scalars at the Large Hadron Collider,''
  Phys.\ Rev.\ D {\bf 79}, 054002 (2009)
  [arXiv:0811.2105]. \\
  T.~Han, I.~Lewis and Z.~Liu,
  ``Colored resonant signals at the LHC: Largest rate and simplest topology,''
  JHEP {\bf 1012}, 085 (2010)
  [arXiv:1010.4309]. \\
  P.~Richardson and D.~Winn,
  ``Simulation of sextet diquark production,''
  Eur.\ Phys.\ J.\ C {\bf 72}, 1862 (2012)
  [arXiv:1108.6154]. \\
  I.~Gogoladze, Y.~Mimura, N.~Okada and Q.~Shafi,
  ``Color triplet diquarks at the LHC,''
  Phys.\ Lett.\ B {\bf 686}, 233 (2010)
  [arXiv:1001.5260]. \\
  I.~Baldes, N.~F.~Bell and R.~R.~Volkas,
  ``Baryon Number Violating Scalar Diquarks at the LHC,''
  Phys.\ Rev.\ D {\bf 84}, 115019 (2011)
  [arXiv:1110.4450]. \\
  Z.~L.~Liu, C.~S.~Li, Y.~Wang, Y.~C.~Zhan and H.~T.~Li,
  ``Transverse momentum resummation for color sextet and antitriplet scalar production at the LHC,''
  Eur.\ Phys.\ J.\ C {\bf 74}, 2771 (2014)
  [arXiv:1307.4341]. \\
  R.~S.~Chivukula, P.~Ittisamai, K.~Mohan and E.~H.~Simmons,
  ``Color discriminant variable and scalar diquarks at the LHC,''    
  Phys.\ Rev.\ D {\bf 92}, no. 7, 075020 (2015)
  [arXiv:1507.06676 [hep-ph]].
  
\bibitem{Bowes:1996xy} 
  J.~P.~Bowes, R.~Foot and R.~R.~Volkas,
  ``Electric charge quantization from gauge invariance of a Lagrangian: A Catalog of baryon number violating scalar interactions,''
  Phys.\ Rev.\ D {\bf 54}, 6936 (1996)
  [hep-ph/9609290]. \\
  E.~Ma, M.~Raidal and U.~Sarkar,
 ``Probing the exotic particle content beyond the standard model,''
  Eur.\ Phys.\ J.\ C {\bf 8}, 301 (1999)
  [hep-ph/9808484]. \\
  K.~S.~Babu, R.~N.~Mohapatra and S.~Nasri,
  ``Post-Sphaleron Baryogenesis,''
  Phys.\ Rev.\ Lett.\  {\bf 97}, 131301 (2006)
  [hep-ph/0606144]. \\ 
  K.~S.~Babu, P.~S.~Bhupal Dev and R.~N.~Mohapatra,
  ``Neutrino mass hierarchy, neutron--anti-neutron oscillation from baryogenesis,''
  Phys.\ Rev.\ D {\bf 79}, 015017 (2009)
  [arXiv:0811.3411 [hep-ph]].
   
\bibitem{Harland-Lang:2014zoa} 
  L.~A.~Harland-Lang, A.~D.~Martin, P.~Motylinski and R.~S.~Thorne,
  ``Parton distributions in the LHC era: MMHT 2014 PDFs,''
  Eur.\ Phys.\ J.\ C {\bf 75}, no. 5, 204 (2015)
  [arXiv:1412.3989].  
   
\bibitem{Dulat:2015mca} 
  S.~Dulat {\it et al.},
  ``New parton distribution functions from a global analysis of quantum chromodynamics,''
  Phys.\ Rev.\ D {\bf 93}, no. 3, 033006 (2016)
  [arXiv:1506.07443]. 
    
\bibitem{Ball:2017nwa} 
  R.~D.~Ball {\it et al.} [NNPDF Collaboration],
  ``Parton distributions from high-precision collider data,''
  Eur.\ Phys.\ J.\ C {\bf 77}, no. 10, 663 (2017)
  [arXiv:1706.00428 [hep-ph]]. 

\bibitem{Mohapatra:1986uf} 
  R.~N.~Mohapatra,
  ``Unification and Supersymmetry: The Frontiers of Quark-lepton Physics,''
  New York, USA: Springer (2003) 421 p.  \\
  Z.~Chacko and R.~N.~Mohapatra,
  ``Supersymmetric $SU(2)_L \times  SU(2)_R  \times  SU(4)_c$ and observable neutron--anti-neutron oscillation,''
  Phys.\ Rev.\ D {\bf 59}, 055004 (1999)
  [hep-ph/9802388]. \\
  B.~Dutta, Y.~Mimura and R.~N.~Mohapatra,
  ``Observable N--anti-N oscillation in high scale seesaw models,''
  Phys.\ Rev.\ Lett.\  {\bf 96}, 061801 (2006)
  [hep-ph/0510291].
  
\bibitem{Arik:2001bc} 
  E.~Arik, O.~Cakir, S.~A.~Cetin and S.~Sultansoy,
  ``A Search for vector diquarks at the CERN LHC,''
  JHEP {\bf 0209}, 024 (2002)
  [hep-ph/0109011]. \\
  H.~Zhang, E.~L.~Berger, Q.~H.~Cao, C.~R.~Chen and G.~Shaughnessy,
  ``Color sextet vector bosons and same-sign top quark pairs at the LHC,''
  Phys.\ Lett.\ B {\bf 696}, 68 (2011)
  [arXiv:1009.5379]. \\
  K.~Das, S.~Majhi, S.~K.~Rai and A.~Shivaji,
  ``NLO QCD corrections to the resonant vector diquark production at the LHC,''
  JHEP {\bf 1510}, 122 (2015)
  [arXiv:1505.07256].  \\
  N.~Assad, B.~Fornal and B.~Grinstein,
  ``Baryon number and lepton universality violation in leptoquark and diquark models,''
  Phys.\ Lett.\ B {\bf 777}, 324 (2018)
  [arXiv:1708.06350 [hep-ph]].

\bibitem{Dobrescu:2018psr} 
  B.~A.~Dobrescu, R.~M.~Harris and J.~Isaacson,
  ``Ultraheavy resonances at the LHC: beyond the QCD background,''
  arXiv:1810.09429 [hep-ph].
  
\bibitem{Hill:2019ldq} 
  C.~T.~Hill, P.~A.~N.~Machado, A.~E.~Thomsen and J.~Turner,
  ``Scalar Democracy,''
  Phys.\ Rev.\ D {\bf 100}, no. 1, 015015 (2019)
  [arXiv:1902.07214]; 
  ``Where are the next Higgs bosons?,''
  Phys.\ Rev.\ D {\bf 100}, no. 1, 015051 (2019)
  [arXiv:1904.04257]. 
    
\bibitem{Sirunyan:2019vgj} 
  A.~M.~Sirunyan {\it et al.} [CMS Collaboration],
  ``Search for high mass dijet resonances with a new background prediction method in proton-proton collisions at $\sqrt{s}=$ 13 TeV,''
  arXiv:1911.03947 [hep-ex]. \\
  CMS Collaboration, ``Searches for dijet resonances in pp collisions at
$\sqrt{s} = 13$ TeV using the 2016 and 2017 datasets'', report CMS-PAS-EXO-17-026
(Sept. 2018), 
\url{https://inspirehep.net/record/1693731}

\bibitem{Sirunyan:2018rlj} 
  A.~M.~Sirunyan {\it et al.} [CMS Collaboration],
  ``Search for pair-produced resonances decaying to quark pairs in proton-proton collisions at $\sqrt{s}=$ 13 TeV,''
  Phys.\ Rev.\ D {\bf 98}, no. 11, 112014 (2018)
  [arXiv:1808.03124 [hep-ex]]. \\
  M.~Aaboud {\it et al.} [ATLAS Collaboration],
  ``A search for pair-produced resonances in four-jet final states at $\sqrt{s} =$ 13 TeV with the ATLAS detector,''
  Eur.\ Phys.\ J.\ C {\bf 78}, no. 3, 250 (2018)
  [arXiv:1710.07171 [hep-ex]].
  
\bibitem{Han:2003wu} 
  T.~Han, H.~E.~Logan, B.~McElrath and L.~T.~Wang,
  ``Phenomenology of the little Higgs model,''
  Phys.\ Rev.\ D {\bf 67}, 095004 (2003)
  [hep-ph/0301040]. \\
  B.~A.~Dobrescu, K.~Kong and R.~Mahbubani,
  ``Prospects for top-prime quark discovery at the Tevatron,''
  JHEP {\bf 0906}, 001 (2009)
  [arXiv:0902.0792 [hep-ph]].

\bibitem{Sirunyan:2019sza} 
  A.~M.~Sirunyan {\it et al.} [CMS Collaboration],
  ``Search for pair production of vectorlike quarks in the fully hadronic final state,''
  Phys.\ Rev.\ D {\bf 100}, no. 7, 072001 (2019)
  [arXiv:1906.11903]; 
  ``Search for vector-like quarks in events with two oppositely charged leptons and jets in proton-proton collisions at $\sqrt{s} =$ 13 TeV,''
  Eur.\ Phys.\ J.\ C {\bf 79}, no. 4, 364 (2019)
  [arXiv:1812.09768 [hep-ex]].  

\bibitem{Aaboud:2018wxv} 
  M.~Aaboud {\it et al.} [ATLAS Collaboration],
  ``Search for pair production of heavy vector-like quarks decaying into hadronic final states in $pp$ collisions at $\sqrt{s} = 13$ TeV,''
  Phys.\ Rev.\ D {\bf 98}, no. 9, 092005 (2018)
  [arXiv:1808.01771];
  ``Combination of the searches for pair-produced vector-like partners of the third-generation quarks at $\sqrt{s} =$ 13 TeV,''
  Phys.\ Rev.\ Lett.\  {\bf 121}, no. 21, 211801 (2018)
  [arXiv:1808.02343].
  
\bibitem{Cheng:2013qwa} 
  H.~C.~Cheng, B.~A.~Dobrescu and J.~Gu,
  ``Higgs mass from compositeness at a multi-TeV scale,''
  JHEP {\bf 1408}, 095 (2014)
  [arXiv:1311.5928].

\bibitem{Dobrescu:2016pda} 
  B.~A.~Dobrescu and F.~Yu,
  ``Exotic signals of vectorlike quarks,''
  J.\ Phys.\ G {\bf 45}, no. 8, 08LT01 (2018)
  [arXiv:1612.01909 [hep-ph]].
    
\bibitem{Luhn:2007yr} 
  C.~Luhn, S.~Nasri and P.~Ramond,
  ``Simple finite non-Abelian flavor groups,''
  J.\ Math.\ Phys.\  {\bf 48}, 123519 (2007)
  [arXiv:0709.1447 [hep-th]]. \\
  T.~Han, I.~Lewis and T.~McElmurry,
  ``QCD corrections to scalar diquark production at hadron colliders,''
  JHEP {\bf 1001}, 123 (2010)
  [arXiv:0909.2666 [hep-ph]].
  
\bibitem{Fortes:2013dba} 
  E.~C.~F.~S.~Fortes, K.~S.~Babu and R.~N.~Mohapatra,
  ``Flavor physics constraints on TeV scale color sextet scalars,''
  arXiv:1311.4101 [hep-ph].
  
\bibitem{Arnold:2012sd} 
  J.~M.~Arnold, B.~Fornal and M.~B.~Wise,
  ``Simplified models with baryon number violation but no proton decay,''
  Phys.\ Rev.\ D {\bf 87}, 075004 (2013)
  [arXiv:1212.4556].

\bibitem{Litos:2014fxa} 
  M.~Litos {\it et al.} [Super-Kamiokande Collaboration],
  ``Search for dinucleon decay into kaons in Super-Kamiokande,''
  Phys.\ Rev.\ Lett.\  {\bf 112}, no. 13, 131803 (2014).

\bibitem{Goity:1994dq} 
  J.~L.~Goity and M.~Sher,
  ``Bounds on $\Delta B = 1$ couplings in the supersymmetric standard model,''
  Phys.\ Lett.\ B {\bf 346}, 69 (1995)
  Erratum: [Phys.\ Lett.\ B {\bf 385}, 500 (1996)]
  [hep-ph/9412208].

\end{thebibliography}
\end{document}